\documentclass[aps,prl,showpacs,amsmath,amsfonts,
               superscriptaddress,twocolumn]{revtex4}
\usepackage{graphicx}
\usepackage{bm}
\usepackage{pslatex}
\newcommand{\eqdef}[1]{\label{eq:#1}}
\renewcommand{\eqref}[1]{Eq.(\ref{eq:#1})}
\newcommand{\br}{{\bf r}}
\newcommand{\bR}{{\bf R}}
\newcommand{\bk}{{\bf k}}

\begin{document}
\title{Consequences of the Pauli exclusion principle for the
        Bose-Einstein condensation of atoms and excitons}

\author{S.~M.~A.~Rombouts, L.~Pollet, K.~Van~Houcke,}
\address{Universiteit Gent - UGent, Vakgroep Subatomaire en Stralingsfysica
         \\
         Proeftuinstraat 86, B-9000 Gent, Belgium
         }
\date{\today}

\begin{abstract}
The bosonic atoms used in present day experiments on
Bose-Einstein condensation are made up of fermionic
electrons and nucleons. In this Letter we demonstrate 
how the Pauli exclusion principle for these constituents 
puts an upper limit on the Bose-Einstein-condensed fraction.
Detailed numerical results are presented for hydrogen atoms
in a cubic volume and for excitons in semiconductors
and semiconductor bilayer systems.
The resulting condensate depletion scales differently
from what one expects for bosons with a repulsive hard-core
interaction. At high densities, Pauli exclusion results in
significantly more condensate depletion.
These results also shed a new light on the 
low condensed fraction in liquid helium II.
\end{abstract}

\maketitle

Recent experiments with ultracold fermionic gases have demonstrated 
the gradual crossover between a Bose-Einstein condensate 
of two-fermion molecules and a BCS-like condensate 
of fermion pairs~\cite{Joch03,Grei03,Zwie03}.
Turning this picture around, one might ask to what extent 
subatomic degrees of freedom play a role 
in Bose-Einstein condensates of bosonic atoms, 
because these atoms are made up of fermions: electrons and nucleons.
From the energetic point of view there is no effect: 
subatomic excitation energies greatly exceed the thermal energy scale 
of Bose-Einstein condensation.
Therefore one can safely assume that the subatomic degrees of freedom 
are completely frozen~\cite{Kett99}.
However, even for a frozen internal structure one has to take into account
the correct symmetries: at the level of the many-electron wave function,
quantum mechanics dictates antisymmetry, 
which makes that electrons can not overlap. 
As a consequence, the Pauli principle for the electrons limits the available 
phase space for the bosonic atoms, which can have an influence
on the properties of the condensate~\cite{Gira63,Gira76,Comb01,Romb03,Avan03,Schm04,Law05}.
It has been demonstrated before 
that a condensate of bosons made up of fermions 
has a maximum occupation number~\cite{Romb02}. 
For hydrogen atoms, that number corresponds to a condensate density 
of the order of $1/(4 \pi a_0^3)$, with $a_0$ the Bohr radius.
Such high densities are not reached in present day experiments
on Bose-Einstein condensates~\cite{Peth02}. 
Still, the Pauli principle can have an effect also at lower densities,
where it leads to condensate depletion.
It is generally believed that it is sufficient to model
this effect through an effective interaction for the bosons 
which is strongly repulsive at short distances,
like a hard-sphere potential 
or e.g. the (unphysical) $r^{-12}$ term in the Lennard-Jones potential.
The condensate will be depleted, simply because of the excluded volume.
However, the only physical parameter which determines the
low-density properties of the condensate is the scattering length.
It is demonstrated below that any bosonic interaction
with the right scattering length 
fails to reproduce the Pauli exclusion effect at high densities.
We show how Pauli exclusion puts an upper bound 
on the Bose-Einstein condensed fraction
of ultracold atomic gases.
The bound is made quantitative for hydrogen atoms,
through the use of an exactly solvable pairing model.
The consequences for ultracold alkali gases, 
exciton condensates in semiconductors
and liquid helium II are discussed.


Following Penrose and Onsager~\cite{Penr56}, one can define 
a Bose-Einstein condensate by looking at the one-boson
density matrix $\rho_B(\br,\br')$ of a many-boson system,
\begin{equation}
  \rho_{B}(\br,\br') = \langle \Phi_B | b^\dag_{\br'} b_\br | \Phi_B \rangle,
  \eqdef{rho_boson}
\end{equation}
with $b^\dag_\br$ the operator that creates a boson
at position $\br$ and $\Phi_B $ the many-boson wave function.
The system is said to exhibit Bose-Einstein condensation
if the one-body density matrix has an eigenvector $\psi_B(\br)$ 
with an eigenvalue $\lambda_B$ of the same order as the total
number of bosons, $N$.
The ratio $f_B=\lambda_B/N$ gives the condensed fraction,
and the eigenvector $\psi_B(\br)$ corresponds 
to the order parameter of the condensate.

Taking into account that atomic bosons are actually
made up of fermions, one realizes that the many-boson state $|\Phi_B \rangle$
corresponds at a more microscopic level 
to a many-fermion wave function $|\Phi_F \rangle$.
For bosons made up of two fermions,
the bosonic one-body density matrix of \eqref{rho_boson}\
can be related to the fermionic two-body density matrix.
Grouping the fermion pair states in a single coordinate $\bR=(\br_1,\br_2)$,
one can write the fermionic two-body density matrix as a square matrix 
$\rho_{F}(\bR,\bR')$.
A Bose-Einstein condensate would show up as an eigenvector $\psi_F(\bR)$ 
of the fermionic two-body density matrix
with a macroscopic eigenvalue $\lambda_F$~\cite{Yang62}.
If the bosons correspond to strongly bound pairs of fermions,
one can expect the bosonic and the fermionic picture of the condensate
to be equivalent~\cite{Gira63}, with
\begin{eqnarray}
  \int \psi_B(\br)  b^\dag_\br d\br & \equiv & \int \psi_F(\br_1,\br_2) 
                              a^\dag_{\br_1} a^\dag_{\br_2} d\br_1 d\br_2, 
  \eqdef{equivalence} \\
  \lambda_B & = & \lambda_F.
\end{eqnarray}
The fermionic model 
has the Pauli correlations between the fermions taken into account,
while the bosonic model does not.
Unfortunately, for any realistic model the fermionic many-body problem
is too complicated in order to determine $\rho_F(\bR,\bR')$ accurately.

A variational approach is feasible:
$\lambda_F$ is an eigenvalue of the many-body operator $B^\dag B$,
with the operator $B^\dag$ defined as
\begin{equation}
  B^\dag = \int \psi_F(\br_1,\br_2) a^\dag_{\br_2} a^\dag_{\br_1} d\br_1 d\br_2.
\end{equation}
If one knows the structure of the order parameter $\psi_F$ 
of the fermionic pair condensate,
one can determine an upper limit for the eigenvalue $\lambda_F$:
\begin{equation}
  \lambda_F =  \langle \Phi_F |  B^\dag B | \Phi_F \rangle 
            <  |E_{P}|,
\end{equation}
where $E_P$ is the ground state energy of a fermionic pairing Hamiltonian,
$ 
  H_P = -B^\dag B .
$
This Hamiltonian is not meant to be phenomenological,
but it is useful here because it is integrable~\cite{Orti04}
and exactly solvable~\cite{Pan98}.
Therefore, given the pair function $\psi_F(\br)$,
one can determine the ground state energy $E_P(2N)$ of $H_P$ for $2N$ fermions, 
and make a rigorous variational statement about the boson condensed fraction:
\begin{equation}
  f_B \leq \frac{|E_P(2N)|}{N}.
  \eqdef{nbvar}
\end{equation}
This upper bound is valid at all temperatures,
but will be most stringent at zero temperature, 
because there one expects the largest condensed fraction.
The resulting eigenstate is not meant as an approximation 
to the true state $| \Phi_f \rangle$, 
because \eqref{nbvar}\ is a variational statement on the condensed fraction,
not on the energy.

To determine the upper bound of \eqref{nbvar},
one needs to know the bosonic order parameter
in terms of the fermionic degrees of freedom.
Although the following treatment is based on a hydrogen 1s wave function,
the procedure is more general.
Alkali atoms such as ${}^{7}$Li, ${}^{23}$Na  or ${}^{87}$Rb can be 
seen as a paired state of a valence electron and a singly ionized core atom.
Because the tail of the wave function for the valence electron is
similar to the hydrogen 1s wave function,
we expect qualitatively the same Pauli effects as for hydrogen atoms.
The Bloch-Messiah theorem says that it is always possible to
write a fermion pair wave function in the form
$  B^\dag =  \sum_i \psi_F(i) a^\dag_{i} a^\dag_{-i},$
where the index $i$ does not necessarily refer to momentum states.
The ground state energy of the corresponding 
pairing Hamiltonian $H_P=-B^\dag B$ 
gives the upper bound for the condensed fraction.
In the low-density limit this results in 
$  f_B \leq 1 - N \sum_i |\psi_F(i)|^4,$
while the upper bound in the high-density limit 
follows from the absolute maximum occupation number 
for the pair condensate~\cite{Romb02}:
$  f_B \leq \frac{\left(\sum_i  |\psi_F(i)|\right)^2}{4N}.$
At intermediate densities one can evaluate the
exact solution for the ground state energy of $H_P$ (see below).

Here we consider hydrogen atoms in a cubic volume $V$ with
periodic boundaries,
because for this case the wave function is known analytically,
and therefore we can determine the upper bound of \eqref{nbvar}\ easily.
At low temperatures and densities, 
one can expect that the protons and electrons form hydrogen atoms 
and that all atoms are in a 1s state.
Because of translational invariance, the bosonic condensate
order parameter will be a uniform function in the center-of-mass
coordinate of the atoms.
The resulting pair operator can be written as
\begin{equation}
  B^\dag = \sum_\bk \psi_F(\bk )a^\dag_{\bk,p} a^\dag_{-\bk,e},
 \eqdef{hydrogen_pairs}
\end{equation}
where the sum runs over all momentum states
allowed by the periodic boundary conditions of the cubic volume $V$,
with the subscripts $p$ and $e$ distinguishing between protons and electrons
and $\psi_F(\bk)$ the pair wave function in momentum space,
$ \psi_F(\bk) = \sqrt{Z}/(1+a_0^2 k^2)^2, $
with $Z=64 \pi a_0^3/V$.
As discussed above, one can assume the internal structure
of the hydrogen atoms to be completely frozen at 
temperatures of the order of $1 \mu K$,
relevant for Bose-Einstein condensation~\cite{Kett99}.
Therefore we can rely on the pair operator of \eqref{hydrogen_pairs}\
in order to evaluate the upper bound of \eqref{nbvar}.

To find the ground-state energy of $H_P$ for $N$ pairs,
one has to solve the following set of equations~\cite{Pan98}:
\begin{equation}
 \frac{1}{y_i} + \sum_{j\neq i} \frac{1}{y_i-y_j}
          = F(2Z y_i),
 \ \ \ \ \forall i=1,\ldots,N-1,
\end{equation}
with the function $F(x)$ given by
\begin{equation}
 F(x) = \frac{32}{\pi} \int_{0}^{+\infty} 
                       \frac{q^2 dq}{x -  \left(1+ q^2\right)^4}.
\end{equation}
The energy $E_P$ is given by
$ E_P = \sum_{i=1}^{N-1} \frac{1}{y_i} - 1.$  
The low density limit is obtained 
by taking the limit $Z \rightarrow 0$ for a fixed number of pairs $N$. 
One obtains, for $n_B a_0^3 \ll 1$, that
 \begin{equation}
  f_B \leq 1 - \frac{33 \pi}{2} ( n_B \, a_0^3).
  \eqdef{nbvar0}
\end{equation}
The result of \eqref{nbvar0}\ was obtained without reference to
interactions. At the bosonic level the Pauli blocking between
the constituting fermions results in a repulsive
interaction between atoms at short distances.
This interaction can be modeled using a pseudopotential 
\begin{equation}
V(r)=\frac{4 \pi \hbar^2 a_s}{m}  \delta(r) \frac{\partial}{\partial r} r,
\end{equation}
with $m$ the atomic mass.
For non-interacting hard spheres, one can identify the scattering
length $a_s$ with the radius~\cite{Huan57,Lee57}.
The Bogoliubov approximation, which applies in the low-density limit,
results at zero temperature in a condensed fraction 
that scales as~\cite{Bogo47,Lee57}
\begin{equation}
  f_B = 1 - \frac{8 }{3\sqrt{\pi}} \left( n_B \, a_s^3 \right)^{\frac{1}{2}}.
  \eqdef{nb_bogol}
\end{equation}
The scaling with a power $1/2$ assures that the variational
bound \eqref{nbvar0}\ is fulfilled.
However, one observes that the Bogoliubov result is fundamentally 
different from  the expression of \eqref{nbvar0}\ 
because it scales differently.
This can be explained by the fact 
that the ground state wave function is not uniform.
In fact, one can expect a higher amplitude for configurations
where the bosons are well separated than for configurations
where bosons nearly overlap~\cite{Feyn53}. 
Consequently, it turns out that Pauli blocking is only a second order effect 
in the low-density limit.
The leading order is determined by bosonic many-body physics,
which not only tries to avoid overlapping atoms,
but also tries to minimize the energy.

\begin{figure}[h]
\begin{center}
\includegraphics[width=8.6cm]{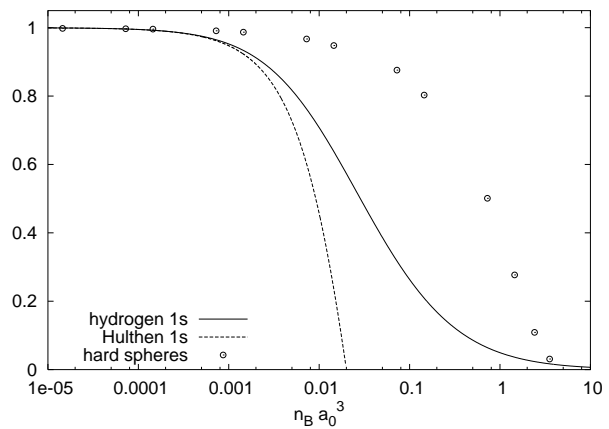}
\caption{Maximal condensed fraction for hydrogen 1s bosons (full line),
 Hulth\'{e}n 1s bosons (dashed line), 
 and bosonic hard spheres of radius $a_s$ (circles),
 as a function of the density parameter $n_B a_0^3$.
 The hydrogen and Hulth\'{e}n curves were calculated for 1000 bosons,
 the hard-sphere results were taken from Ref.\cite{Gior99}.
\label{fig:maxcf}}
\end{center}
\end{figure}

The high density limit can be derived from the maximal occupation
number of the pair state~\cite{Romb02},
which for hydrogen results in
$ f_B \leq \frac{1}{4 \pi n_B \, a_0^3}.$
%
%
At high densities one has to assess the effects
of the interactions with the other atoms.
Calculations with a quantum Monte Carlo method~\cite{Gior99} 
and a constrained variational method~\cite{Cowe02}
have shown that for an interaction with scattering length $a_s$ 
the condensed fraction is well described 
by the Bogoliubov approximation of \eqref{nb_bogol},
up to densities of the order of $n_B \sim 10^{-3} a_s^{-3}$.
For hydrogen atoms the s-wave scattering length 
is given by $a_s=0.41 a_0$~\cite{Jami92}.
One can see in Fig.~\ref{fig:maxcf} that at a density
of $n_B = 10^{-3} a_s^{-3} = 1.5 \times 10^{-2}a_0^{-3}$,
the Pauli effect results in a condensate depletion of the order of $36\%$, 
while hard spheres with a radius $a_s$,
and hence any interaction with the same scattering length,
yield a depletion of only $5\%$.
One can conclude that at densities of the order
of $10^{-3} a_0^{-3}$ and higher 
the Pauli exclusion effect results in a significantly
stronger condensate depletion than the hard-sphere potential,
and that an effective two-body interaction for the bosons is not able 
to reproduce this effect.

However, at high densities one also has to consider another effect:
due to interactions with electrons and protons in neighboring atoms,
the intrinsic wave function of the atoms might get deformed
from the standard hydrogen 1s wave function.
A way to take these interactions into account 
is to use a screened Coulomb potential, 
such as the Hulth\'{e}n potential~\cite{Hult42,Bany86}:
$ V_H(r)= -\frac{4 \gamma V_0}{e^{2 \gamma r/a_0}-1}, $
with $V_0$ the hydrogen $1s$ binding energy
and $\gamma$ a dimensionless parameter
proportional to the screening constant as defined
in the Debeye-H\"{u}ckel or Thomas-Fermi models.
This is a phenomenological way to take screening into account,
which could deviate from the true microscopic behavior at high densities.
The intrinsic wave function in momentum space becomes
$  \psi_F(\bk) = \sqrt{Z \left(1-\gamma^2\right)}/
                 \left[\left((1-\gamma)^2+a_0^2 k^2\right) 
                 \left((1+\gamma)^2+a_0^2 k^2\right)\right]. $
In theory, at very high densities the system could undergo a Mott transition,
where electrons and protons are no longer bound to each other in hydrogen atoms,
but rather form a plasma. 
For cold hydrogen atoms this is not a realistic scenario,
but it does apply to excitons in semiconductors,
which have a similar intrinsic structure.
The Mott transition occurs at an exciton density of $n_c \simeq 0.02 a_x^{-3}$,
with $a_x$ the exciton Bohr radius~\cite{Mott90}.
This behavior should be reflected in the density dependence of the 
screening parameter $\gamma$:
at very low densities $\gamma$ should tend to zero,
in order to recover the hydrogen wave function from
the Hulth\'{e}n wave function.
The Mott transition, on the other hand,  would require
the bound states of the Hulth\'{e}n potential 
to disappear around the transition density. 
As this happens at $\gamma=1$, we follow Ref.~\cite{Tang02} to take
the screening parameter $ \gamma = \sqrt{n_B/n_c}. $
The variational bound for the condensed fraction 
can be obtained for any density by solving the eigenvalue 
equations of the pairing Hamiltonian $H_P$
with the screened pair structure.
Fig.~\ref{fig:maxcf} shows the results
for the hydrogen 1s and Hulth\'{e}n 1s intrinsic wave functions.
At the resolution of the figure the results for 1000 and 2000
particles are indistinguishable, meaning 
that convergence has been reached and 
that finite-size effects are negligible at these particle numbers. 
The screening effects 
enhance the condensate depletion even more. They
become important only when the maximal
condensed fraction is lower than 80\%,
so they are of secondary importance in the high density regime.

\begin{figure}[h]
\begin{center}
\includegraphics[width=8.6cm]{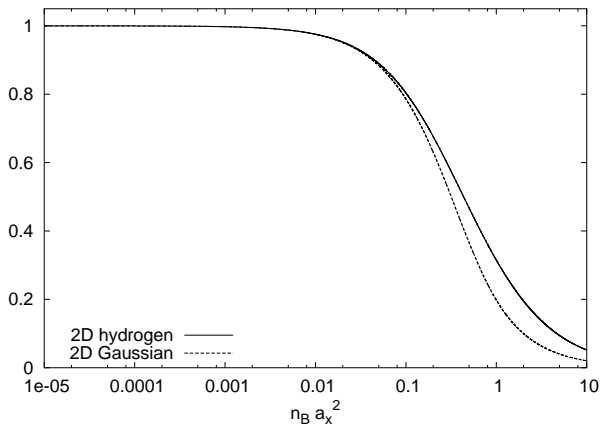}
\caption{Maximal condensed fraction for 
 two-dimensional excitons
 with a hydrogen-like (solid line) or Gaussian (dashed line) intrinsic wave 
 function,  as a function of the exciton density parameter $n_b a_x^2$.
 The curves were calculated for 1000 excitons.
\label{fig:maxcf2}}
\end{center}
\end{figure}

The densities where the Pauli effect becomes sizeable
are probably out of reach for an ultracold atomic hydrogen gas,
because three-body recombination processes would convert the atoms 
into molecules.
However, the same pair structure also applies
to Wannier excitons in semiconductors.
There the Pauli effect might explain, together with
biexciton recombination, why a clear signal of 
exciton condensation has not yet been observed
in a three-dimensional structure.
Bose-Einstein condensation of excitons has been observed
in semiconducting bilayers~\cite{Eise04}.
The physics there is basically two-dimensional~\cite{Butov02,Lai04}.
There too the Pauli effect applies.
Although an analytical expression for the intrinsic structure
of these excitons is not readily available,
one can estimate the Pauli effect by looking at the 
results of Fig.~\ref{fig:maxcf2} for a two-dimensional hydrogen wave function,
$ \psi_{F}(k) \propto (1+a_x^2 k^2/4)^{-3/2} $
and for a Gaussian wave function with the same low-density properties,
$ \psi_{F}(k) \propto \exp(-a_x^2 k^2/5) $, with $a_x$ the two-dimensional
excitonic Bohr radius.


If one has to treat more valence electrons independently,
then the boson operator corresponds to a three- or higher-body 
fermion operator instead of a pair operator,
and the resulting Hamiltonian $-B^\dag B$ is no longer exactly solvable.
Still, one can expect the Pauli principle to have
qualitatively the same effect on e.g. ${}^4$He condensates:
the electrons will avoid overlap and hence limit the available
phase space for the ${}^4$He atoms.
A uniform distribution of hard-core bosons can qualitatively explain 
the reduced condensed fraction in liquid helium II 
compared to an ultracold low-density Bose gas~\cite{Feyn53,Penr56}.
The Pauli blocking of the underlying fermions offers a
more microscopic view of this process.
Pauli effects will result in a significant depletion of the condensate 
at densities of the order of $10^{-3}$ times the close-packing density
or higher, and definitely at the density of liquid helium.

We have demonstrated here that Pauli blocking of the underlying electrons
leads to condensate depletion in ultracold atomic gases
and in exciton condensates.
This effect might be measurable in systems 
where densities of the order of $10^{-3} a_0^{-3}$ can be reached, 
such as Wannier excitons in semiconductors
or ${}^4$He films adsorbed on porous Vycor glass~\cite{Croo83}.
This effect depends solely on the symmetry and
internal structure of the wave-functions.
In the high-density regime, this effect can not be modeled
through an effective two-body interaction at the bosonic level.
Interactions might change the internal wave function of the fermionic pairs. 
Our calculations based on a screened potential show 
that the Pauli effect dominates over the interaction effects
at densities one or two orders of magnitude below the Mott transition density.

%
The authors wish to thank D.~Van~Neck, K.~Heyde and J.~Dukelsky 
for interesting discussions and suggestions, 
S.~Giorgini for providing details of the hard-sphere calculations,
and the Fund for Scientific Research - Flanders (Belgium) for financial support.


\end{document}